\documentclass[]{article}

\usepackage[utf8]{inputenc}
\usepackage{hyperref}
\usepackage{url}
\usepackage{fullpage}
\usepackage{subfigure}
\usepackage{amsmath}
\usepackage{amssymb}
\usepackage{latexsym}
\usepackage{amsthm}
\usepackage{boxedminipage}
\usepackage{listings}
\usepackage{ifpdf}

\ifpdf
\usepackage[pdftex]{graphicx}
\else
\usepackage{graphicx}
\fi

\newtheorem{definition}{Definition}
\newtheorem{example}{Example}
\newtheorem{remark}{Remark}

\title{An Array Algebra}
\author{Albrecht Schmidt\footnote{European Space Astronomy Centre, Villanueva de la Cañada, Madrid, Spain}}

\date{2008-11-28}

\begin{document}

\ifpdf
\DeclareGraphicsExtensions{.pdf, .jpg, .tif}
\else
\DeclareGraphicsExtensions{.eps, .jpg}
\fi

\maketitle

\begin{abstract}
    This is a proposal of an algebra which aims at distributed array processing.  The focus lies on re-arranging and distributing array data, which may be multi-dimensional.  The context of the work is scientific processing; thus, the core science operations are assumed to be taken care of in external libraries or languages.  A main design driver is the desire to carry over some of the strategies of the relational algebra into the array domain.
\end{abstract}

\section{Introduction}

Algebras have been used to describe operations common in data management for their conciseness and well-definedness~\cite{AHV95}.  The underlying collection models for these algebras are usually sets or multi-sets of tuples.  In scientific processing, however, many of the interesting data are array-type and abound so that they lend themselves to distributed storing and processing.  This note tries to explore the implications of the three technical challenges by re-phrasing some core ideas of the well-known relational algebra in an distributed array-context.  Hopefully, opportunities for combining the two worlds show up.

\subsection{Design Considerations} % (fold)
\label{sub:design_considerations}

One reason for staring out from relational processing~\cite{AHV95} is that many large-scale and transaction-safe systems have been built with great success on this paradigm.  To make best use of the information workers' skill-base, it would be interesting to see how to transition smoothly between the relational and the array paradigms.  Therefore, this note will first explore how to re-phrase the relational operations of selection, projection, union, and join in an array context.

From a training and knowledge standpoint, it is desirable to re-use as many concepts from existing technologies as possible.  We therefore try to adapt well-understood concepts from data processing and carry them over into the related domain of array processing.  These concepts also have the advantage that they are conceptually simple but still very useful and relevant.

It is desirable to use algebraic operations since they are not only useful for extracting information from data but also for distributing data in a shared-nothing environment.  Thanks to the closure properties of the algebra, it is possible to use same language for querying the data that is used for distributing them.  Eventually, a distributed representation of the input data is just an algebraic view on the input data and could be used straight-away in queries.

% subsection design_considerations (end)

\subsection{Data Modelling Aspects} % (fold)
\label{sub:data_modelling_aspects}

In many programming languages, arrays are usually of fixed dimension, just like matrices in mathematics.  For long-lived data that reside in databases, it might be desirable to lift this restriction and allow more flexibility.  After all,  append is an operation that is very common in data management.

Non-fixed dimensionality does not necessarily imply that the data is a time series although this is often possible or a matter of interpretation.  Data may be appended for a variety of reasons, including bulk-loading of data and recording of asynchronous events.

% subsection data_modelling_aspects (end)

\section{The Algebra}

The building blocks of the algebra are functions which map arrays to arrays.  In the sequel, we present the function we consider useful for dealing with the range of problems laid out in the previous section.

\subsection{Basic Definitions} % (fold)
\label{sub:basic_definitions}

For the purpose of this paper, we interpret an array as a function from an index into a domain.  In the case of a vector $[3.1,3.14,3.141,3.1415]$ the index might be the number 0 to 3, and the domain might be the real numbers.  Arrays suggest that the primary way of accessing the contents is by going through the index.  While this might be true for numerical application, other applications and data types might require a wider range of access paths such as scans, indexes, and sub-arrays.

\begin{definition}
    An $n$-dimensional array $A$ is a function $A:I\to(D\cup\bot)$, where $I\subseteq I_1\times I_2\dots\times I_n$ is an $n$-dimensional set of indices and $D$ is the domain of the array entries; $\bot$ denotes that the function is undefined at certain indices.  We use the notation $i\to d$ to denote to which $d\in(D\cup\bot)$ an $i\in I$ is mapped; as a shorthand to indicate what array we are operating on, we also write $A(i)=d$.
\end{definition}

\begin{remark}
    $D$ may an arbitrarily complex domain including composite data types and, recursively, arrays.  It also stands for the basic building blocks such as integers, floats, and strings.
\end{remark}

\begin{example}
    The matrix 
    $M=\left( \begin{array}{cc}
    a & b \\
    c & d \end{array} \right)$
    could be represented as a two-dimensional array as follows:  $I=\{(0,0),(0,1),(1,0),(1,1)\}$, $A:\{(0,0)\to a,(0,1)\to b,(1,0)\to c,(1,1)\to d\}$, and $D=\{ a,b,c,d \}$.  For individual elements, we also write $M(1,1)=d$.
\end{example}

\begin{definition}
    The space of all arrays is denoted $\mathbb{A}$, the space of all indices is $\mathbb{I}$, and the space of all domains $\mathbb{D}$.
\end{definition}

Using these definitions, an array database could be seen as:\[
    \mathrm{db}\subseteq\mathbb{A}
\]

Using these ideas, we proceed to the definition of useful functions to work with arrays.
% subsection basic_definitions (end)

\subsection{Functions} % (fold)
\label{sub:functions}

We now define some of the basic function we need to work with arrays.  Most of them are inspired by the relational algebra and try to carry over its simplicity into the array domain.

We start out by defining the array equivalent of relational projection.

\begin{definition}
    The array projection function $\pi_{J}:A\to B$, $A:I\to D, J\subseteq I$ and $B$ are arrays such that $\pi_J(A)=\{a\to b\;|\;a\in A(I\cap J), b=A(a) \}$.
\end{definition}

\begin{remark}
    We require $J\subseteq I$ although this is not strictly necessary.  Furthermore, note that projection does not change the dimensions of the array.  To change them, we require an index transformation, possibly following a projection.
\end{remark}

The design decision to separate projection, \emph{i.e.}, filtering on the index set, and dimensionality reduction is the desire to keep the basic operations orthogonal.

\begin{example}
    To extract the first column of the matrix in the previous example, we write $\pi_{\{(0,0),(1,0)\}}(A)=\{(0,0)\to a,(0,1)\to b\}$.
\end{example}

\begin{definition}
    The array selection function $\sigma_c:A\to B, A:I\to D$ is defined as follows: $\sigma_c(I\to D)=\{ i\to d\;|\;i\in D, d\in D, A(i)=d, c(d)\;\text{holds}\}$.
\end{definition}

Like in the relational case, a selection does not change the `schema' of an array; it only filters an array by content.

\begin{example}
    To extract the element $b\in D$ in the example Matrix, we write $\sigma_{=b}(A)=\{(1,0)\to b\}$
\end{example}

\begin{remark}
    The Algebra operations defined so far are also meant to illustrate the difference in data model and querying to relational databases.
\end{remark}

\begin{definition}
    The cross product between two arrays $A_1$ and $A_2$ is defined as follows: $A_1\times A_2=\{a_1\circ a_2 \to d_1\circ d_2\;|\;a_1\to d_1 \in A_1, a_2\to d_2 \in A_2\}$, where $\circ$ denotes concatenation
\end{definition}

Like projection, the result of a cross product has a schema that is different from the input schemas.

\begin{remark}
    If $n_1$ and $n_2$ are the dimensions of $A_1$ and $A_2$, then the dimension of the $A_1\times A_2$ is $n_1+n_2$.  If $s_1$ and $s_2$ are the number of elements in $A_1$ and $A_2$, then $n_1\cdot n_2$ is the number of elements in $A_1\times A_2$.  Like an ordinary cross-product, $\times$ on arrays does not add more information nor does it throw away information.
\end{remark}

To change the index sets of arrays, we now define the notion of \emph{index transformation}.  Such an operation may be desirable for a variety of reasons.  For example, it might be useful to ensure that the range of the index attributes is compacted and sparsity is avoided.  It is also useful to be able to add or remove dimensions to ensure compatibility between arrays.

\begin{definition}
    We define three kinds of \emph{index transformations}.
    An \emph{index augmentation} is a bijective function $f:I\to J$ where $I\subseteq I_1\times\dots\times I_n$ and $J\subseteq I_1\times\dots\times I_n\times\dots\times I_m$.  An \emph{index reduction} is a bijective function $f:I\to J$ where $I\subseteq I_1\times\dots\times I_n\times\dots\times I_m$ and $J\subseteq I_1\times\dots\times I_n$.  An index reorganisation is a bijective function $f:I\to I$ where $I\subseteq I_1\times\dots\times I_n$.
\end{definition}

\begin{remark}
    Index transformations are used to increase the dimensions of an index and move around/translate data in space.  Note that both augmentations and reductions are not allowed to add or drop information.
\end{remark}

Now that we can ensure the compatibility of arrays in terms of dimensionality, we are able to describe the array equivalent of the union operation:

\begin{definition}
    The \emph{union} operation is defined as follows: $A\cup B=\{i\to d\;|\;i\to d\in A\vee i\to d\in B\}$
\end{definition}

\begin{remark}
    Note that the union operation should fail if there is an $i'\to d'$ where $i=i'$ and $d\not=d'$
\end{remark}

In a later section, these basic operations will be used to express tasks of interest.

% subsection functions (end)

\section{Properties} % (fold)
\label{sec:properties}

Here, we discuss some of the modelling capabilities of the algebra by going through common implementation issues.  The goal is to show that the algebra achieves its design goal to be implementation-neutral in many respects, \emph{i.e.}, it does not impose a particular way of implementation.

\subsection{Array Representation} % (fold)
\label{sub:array_representation}

There are many well-known techniques to store arrays in main-memory and secondary memory.  The algebra we present aims not to favour any particular way.  It is designed to be as neutral as possible towards issues such as row-major, column-major, sparse, dense, locality-preserving, centralised, distributed \emph{etc}~storage.

A storage or query engine is even allowed to keep various copies of an association around or break up association as long as consistency is preserved, \emph{i.e.}, for any two association $i\to d$ and $i'\to d'$: $i=i'\implies d=d'$.  

% subsection array_representation (end)

\subsection{Keys and Indices} % (fold)
\label{sub:keys_and_indices}

According to our definitions, numerical indices are the primary access paths to the array contents, \emph{i.e.}, they play the role of keys in relational models.  In practise, however, other types of keys, such as (length-limited) strings, timestamps or enumeration datatypes (also known as attribute names in the relational model), are important as well.

% subsection keys_and_indices (end)

\subsection{Algebraic Properties} % (fold)
\label{sub:algebraic_properties}

The system defined above is an algebra in the sense it defines a set and functions from the set into the set.  The properties of the individual functions are beyond this note but become important when queries are executed.

% subsection algebraic_properties (end)

\subsection{Mixing Relational and Array} % (fold)
\label{sub:mixing_relational_and_array}

Users often would like to mix ideas from the relational world and the array world.  As a simplified example, consider a relational schema like $R(\textit{\underline{measurementID}},\textit{time},\textit{detector},\textit{valueMatrix})$; this is often a good approach to store the value matrices returned by a detector during a measurement.  Note the uniqueness or key constraint on the first column.

In an array world, there are various avenues to storing this kind of information.  One way would be to represented as a (two-dimensional) matrix where one dimension, the `columns', is an enumerated data type $\{\textit{\underline{measurementID}},\textit{time},\textit{detector},\textit{valueMatrix}\}\sim\{0,1,2,3\}$; the other dimension, the `rows', would then be the numbers $[0,..,n]$, where $n$ is the number of measurements.  The $\textit{valueMatrix}:[0,n]\times[0,m]\to\textit{Float}$ is then a conventional Matrix.  Note that this schema is recursive on the type level.

Furthermore, one could make use of the uniqueness constraint on $\textit{measurementID}$ and identify it with the `row' dimension.  Note that we do not argue for a particular way to enforce uniqueness or key properties.  Since we aim at scientific processing where huge data volumes easily require application-tailored algorithms, the enforcement might well be done in an application-specific manner, for example, by devising special ETL mechanisms.  Of course, uniqueness could also be enforced with a traditional index on the key attribute.  However, this is not in every application the most practical method.

% subsection mixing_relational_and_array (end)

% section properties (end)

\section{Common Tasks} % (fold)
\label{sec:common_tasks}

This section presents some common data processing tasks and how are they implemented in the algebra of the previous sections.  We will start by looking at common join operations.

\subsection{Equi-Joins} % (fold)
\label{sub:equi_joins}

In our context, an equi-join is a cross product whose result is filtered on the equality of a predicate.  In an array context, the equality predicate could be applied to the index as well as to the domain value.  However, since the model presented in this paper treats the actual array contents as a black-box, we focus on the equality of the indices, and view an equi-join as an expression of the form \[
    \sigma_p(A\times B).
\]

% subsection equi_joins (end)

\subsection{Semi-Joins} % (fold)
\label{sub:semi_joins}

Similarly, a semi-join is a cross product followed by a projection and selection.  Thus it is of the form: \[ \pi_I(\sigma_p(A\times B)) \]

% subsection semi_joins (end)

\subsection{Anti-Joins} % (fold)
\label{sub:anti_joins}

Similarly, an anti-join is a cross product followed by a selection.  Thus it is of the form: \[ \sigma_p(A\times B) \]

% subsection anti_joins (end)

\subsection{Distributing Data} % (fold)
\label{sub:distributing_data}

Using the algebra operations laid out in this note, we are able to distribute an array by both \emph{vertical} and \emph{horizontal} partitioning.  Vertical partitioning can easily be implemented using the union operation.  Horizontal partitioning is sensible when the data item associated with an index entry is particularly large.  In this case, the data item can be split up into two parts by duplicating the index and associating a part of the entry with every copy of the index.  The separated items can be joined together using equi-joins which, due to the uniqueness of the index, are particularly cheap to execute.

% subsection distributing_data (end)

% section common_tasks (end)

\section{Conclusion} % (fold)
\label{sec:conclusion}

This paper presents an algebra for distributing and querying arrays.  Since array operations tend to be complex and done in an external general purpose programming language, the algebra aims to take after the design principles of the relational algebra.  We presented translations of relational concepts into the array world.  Future work includes the implementation and evaluation of the concepts.

% section conclusion (end)

\section{Acknowledgements} % (fold)
\label{sec:acknowledgements}

The author would like to thank Harold Metselaar for useful discussions.

% section acknowledgements (end)
\bibliographystyle{plain}

\end{document}